# Existence and properties of the Navier-Stokes equations

## A.V. Zhirkin

National Research Centre " Kurchatov Institute ", Kurchatov Centre of Nuclear Technologies, Fusion Reactor Department, Reactor Problem Laboratory, Russian Federation, Moscow 123182, Academician Kurchatov square, 1.
e-mail: aleksej-zhirkin@yandex.ru
e-mail:aleksej-zhirkin@yandex.ru <mailto:aleksej-zhirkin@yandex.ru>

## Abstract

A proof of existence, uniqueness and smoothness of the Navier-Stokes equations is an actual problem, which solution is important for different branches of science. The subject of this study is obtaining the smooth and unique solutions of the three-dimension Stokes-Navier equations for the initial and boundary value problem. The analysis shows that there exist no viscous solutions of the Navier-Stokes equations in three dimensions. The reason is the insufficient capability of the divergence-free velocity field. It is necessary to modify the Navier-Stokes equations for obtaining the desirable solutions. The modified equations describe a three-dimension flow of incompressible fluid which sticks to a body surface. The equation solutions show the resonant blowup of the laminar flow, laminar-turbulent transition, the fluid detachment that opens the way to solve the magnetic dynamo problem.

## About the author(s)

Alexey V. Zhirkin is a Senior Researcher at the Department of Fusion Reactors of the National Research Centre "Kurchatov Institute" in Moscow, Russian Federation. He graduated from the Moscow Engineering Physics Institute (PhD) and finished the Doctor course at the Department of Quantum Engineering and Systems Science in the University of Tokyo, Japan. His research area is neutronics analysis of fusion devices. He is involved in group's research considering the design of the tokamak-based hybrid fusion-fission neutron source, including the problems of plasma confinement, magnetohydrodynamic effects, and thermal hydraulics, etc.

## Public Interest Statement

The Navier-Stokes equations are one of the fundamentals of fluid mechanics. They describe the motion of viscous liquids, gases or plasmas, and are the basis of mathematical modeling of many natural phenomena. Their applications cover virtually all areas of science: aerodynamics and hydrodynamics, astrophysics, engineering, medicine and health, chemistry, biology, geophysics, electro and magnetic hydrodynamics, etc. However, the solutions of the Navier-Stokes equations suitable for analysis are found only for simplified problems that have little or no physical interest. There is no complete understanding of the properties of these equations. An open problem in mathematics is whether their physically reasonable (smooth) solutions always exist in three dimensions. Also it is very important to understand how the Navier-Stokes equations describe the chaotic behavior of fluids (turbulence) that is one of the most difficult problems of mathematical physics. This article is the response to the Navier–Stokes existence and smoothness challenge.

**Main Text**

# 1. Introduction

A proof of existence, uniqueness, and regularity of three-dimensional fluid flows for an incompressible fluid concerns most mathematically complex unsolved problems (Constantin, 2001; Fefferman, 2000; Ladyzhenskaya, 2003). The positive result is obtained in two space dimensions (Ladyzhenskaya, 1969). In three dimensions, existence is proved for weak solutions (Leray, 1934), but their uniqueness and smoothness are not known. For strong solutions, the three-dimensional problem is not solved till now (Constantin, 2001; Fefferman, 2000; Ladyzhenskaya, 2003).

The solution of the problem can help ensure the development in different branches of science. The significance of the challenge is well described in the work (Fefferman, 2000).

"A fundamental problem in analysis is to decide whether such smooth, physically reasonable solutions exist for the Navier–Stokes equations. … Fluids are important and hard to understand. There are many fascinating problems and conjectures about the behavior of solutions of the Euler and Navier–Stokes equations. … Since we don't even know whether these solutions exist, our understanding is at a very primitive level. Standard methods from PDE appear inadequate to settle the problem. Instead, we probably need some deep, new ideas."

In the work of (Fefferman, 2000), the detailed statement for the Cauchy initial value problem and the problem with the initial periodical conditions of the Navier–Stokes equations is presented. We use this statement as the basic formulation of problem-solving in our research.

The regularity question considering the boundary value problem is posed in the work (Constantin, 2001):

"What are the most general conditions for smooth incompressible velocities $\vec{v}(0,\vec{r})$ that ensure, in the absence of external input of energy, that the solutions of the Navier–Stokes equations exist for all positive time and are smooth? … A specific regularity question, still open, is for instance: Given an arbitrary infinitely differentiable, incompressible, compactly supported initial velocity field in $R^3$, does the solution remain smooth for all time? A version of the same question is: Given an arbitrary threedimensional divergence-free periodic real analytic initial velocity, does the ensuing solution remain smooth for all time?"

In the work of (Ladyzhenskaya, 2003), there is a more general formulation of the problem,

Do the Navier–Stokes equations together with the initial and boundary conditions provide the deterministic description of the dynamics of an incompressible fluid or not?

In our research, we will try to find the answers to all these questions.

# 2. Analysis

## 2.1. Statement of problem

The system of the non-stationary Navier-Stokes equations for an incompressible fluid in three dimensions looks as follows:

$$\rho \cdot \partial \vec{v}(t,\vec{r})/\partial t + \rho \cdot (\vec{v}(t,\vec{r}) \cdot \nabla) \cdot \vec{v}(t,\vec{r}) = -\nabla p(t,\vec{r}) + \eta \cdot \Delta \vec{v}(t,\vec{r}) + \vec{f}_{ex}(t,\vec{r}), \qquad (1)$$

$$div(\vec{v}(t,\vec{r})) = 0, \qquad (2)$$

$$(\vec{r} \in R^3, t \geq 0)$$

where $\rho$ is the fluid density, $\vec{v}(t,\vec{r})$ is the vector field of velocities of a fluid at time $t$ and a point with coordinates $\vec{r} = (x, y, z)$, $p(t,\vec{r})$ is the pressure of a fluid, $\eta$ is the dynamic viscosity, $\vec{f}_{ex}(t,\vec{r})$ is the external force defined as a smooth function on $[0,\infty) \times R^3$. $\nabla$ is the Hamilton operator, and $\Delta$ is the Laplace operator. Equation (1) is the Cauchy momentum equation and Equation (2) is the mass continuity equation.

If the initial value problem is considered, the initial condition looks as follows:

$$\vec{v}(0,\vec{r}) = \vec{v}_0(\vec{r}) \tag{3}$$

where $\vec{v}_0(\vec{r})$ is an arbitrary divergence-free vector field defined in $R^3$.

For the boundary value problem (Constantin, 2001; Ladyzhenskaya, 2003), the boundary condition is taken for a fluid sticking to a body surface:

$$\vec{v}(t,\vec{r}_S) = 0 \tag{4}$$

where $S$ is the surface separating the fluid and body.

The Clay Mathematics Institute suggests to prove or deny existence of solutions satisfying (1), (2), and (3) for viscous fluid ($\eta > 0$) under the following conditions (Fefferman, 2000).

For physically reasonable solutions, $\vec{v}(t,\vec{r})$ does not grow large as $|\vec{r}| \to \infty$. Hence, the forces $\vec{f}_{ex}(t,\vec{r})$ and initial conditions $\vec{v}_0(\vec{r})$ satisfy

$$\left| \partial_{\vec{r}}^{\alpha} \vec{v}_0(\vec{r}) \right| \le C_{\alpha,K} \cdot (1 + |\vec{r}|)^{-K} \text{ on } R^3, \text{ for any } \alpha \text{ and } K \tag{5}$$

and

$$\left| \partial_{\vec{r}}^{\alpha} \partial_t^m \vec{f}_{ex}(t,\vec{r}) \right| \le C_{\alpha,m,K} \cdot (1 + |\vec{r}| + t)^{-K} \text{ on } [0,\infty) \times R^3, \text{ for any } \alpha, m \text{ and } K. \tag{6}$$

A solution of (1), (2), and (3) is accepted as physically reasonable only if it satisfies

$$\vec{v}(t,\vec{r}), \, p(t,\vec{r}), \, f_{ex}(t,\vec{r}) \in C^{\infty}([0,\infty) \times R^3), \; \vec{v}_0(\vec{r}) \in C^{\infty} \; (\vec{r} \in R^3), \tag{7}$$

and

$$\int_{R^3} \left| \vec{v}(t,\vec{r}) \right|^2 \cdot d\vec{r} < const \text{ for all } t \ge 0 \text{ (bounded energy)}. \tag{8}$$

**Note.** The function $\vec{f}_{ex}(t,\vec{r})$ is a smooth one on $[0,\infty) \times R^3$. So we added the condition $f_{ex}(t,\vec{r}) \in C^{\infty}([0,\infty) \times R^3)$ in (7). This condition is absent in the same expression of the work of Fefferman (2000).

Alternatively, for spatially periodic solutions of (1), (2), and (3), we assume that $\vec{v}_0(\vec{r})$, $\vec{f}_{ex}(t,\vec{r})$ satisfy

$$\vec{v}_0(\vec{r}) = \vec{v}_0(\vec{r} + \vec{e}_j), \; \vec{f}_{ex}(t,\vec{r}) = \vec{f}_{ex}(t,\vec{r} + \vec{e}_j) \text{ on } [0,\infty) \times R^3, \; j = 1,2,3 \tag{9}$$

where $\vec{e}_j$ is $j$th unit vector in $R^3$.

In place of (5) and (6), we assume that $\vec{v}_0(\vec{r})$ is smooth and that

$$\left| \partial_{\vec{r}}^{\alpha} \partial_t^m \vec{f}_{ex}(t,\vec{r}) \right| \le C_{\alpha,m,K} \cdot (1 + |t|)^{-K} \text{ on } [0,\infty) \times R^3, \text{ for any } \alpha, m \text{ and } K. \tag{10}$$

We accept a solution of (1), (2), or (3) as physically reasonable if it satisfies

$$\vec{v}(t,\vec{r}) = \vec{v}(t,\vec{r} + \vec{e}_j), \; p(t,\vec{r}) = p(t,\vec{r} + \vec{e}_j) \text{ on } R^3, \; j = 1,2,3 \tag{11}$$

and

$$\vec{v}(t,\vec{r}), \, p(t,\vec{r}), \, f_{ex}(t,\vec{r}) \in C^{\infty}([0,\infty) \times R^3). \tag{12}$$

The problem is formulated in four statements (Fefferman, 2000).

**(A) Existence and smoothness of Navier–Stokes solutions on $R^3$.** Take $\eta > 0$ and space dimension $n = 3$. Let $\vec{v}_0(\vec{r})$ be any smooth, divergence-free vector field satisfying (5). Take $\vec{f}_{ex}(t,\vec{r})$ to be identically zero. Then there exist smooth functions $p(t,\vec{r})$, $\vec{v}(t,\vec{r})$ on $[0,\infty) \times R^3$ that satisfy (1), (2), (3), (7), and (8).

**(B) Existence and smoothness of Navier–Stokes solutions in $R^3/Z^3$.** Take $\eta > 0$ and $n = 3$. Let $\vec{v}_0(\vec{r})$ be any smooth, divergence-free vector field satisfying (9); we take $\vec{f}_{ex}(t,\vec{r})$ to be identically zero. Then there exist smooth functions $p(t,\vec{r})$, $\vec{v}(t,\vec{r})$ on $[0,\infty) \times R^3$ that satisfy (1), (2), (3), (11), and (12).

**(C) Breakdown of Navier–Stokes solutions on** $R^3$. Take $\eta > 0$ and $n = 3$. Then there exist a smooth, divergence-free vector field $\vec{v}_0(\vec{r})$ on $R^3$ and a smooth $\vec{f}_{ex}(t,\vec{r})$ on $[0,\infty) \times R^3$, satisfying (5), and (6) on $[0,\infty) \times R^3$, for which there exist no solutions $(p(t,\vec{r}), \vec{v}(t,\vec{r}))$ of (1), (2), (3), (7), and (8) on $[0,\infty) \times R^3$.

**(D) Breakdown of Navier–Stokes Solutions on** $R^3/Z^3$. Take $\eta > 0$ and $n = 3$. Then there exist a smooth, divergence-free vector field $\vec{v}_0(\vec{r})$ on $R^3$ and a smooth $\vec{f}_{ex}(t,\vec{r})$ on $[0,\infty) \times R^3$, satisfying (9), and (10), for which there exist no solutions $(p(t,\vec{r}), \vec{v}(t,\vec{r}))$ of (1), (2), (3), (11), and (12) on $[0,\infty) \times R^3$.

## 2.2. Idea of proof

According to the Cauchy–Kowalevski theorem, for the local existence and uniqueness of the solution of a Cauchy initial value problem for a linear partially differential equation with constant coefficients, it is enough to prove the expansibility of this solution to a uniformly converging power series in some neighborhood of every point in the solution domain. The solution can be a function of real or complex variables (Courant & Hilbert, 1962; Vladimirov, 1983).

We will extend this statement to the nonlinear Navier-Stokes equations and generalize the result obtained for a local solution to all space $R_+^1 \times R^3$ to obtain a global solution. We can do it by the following way.

At first, we carry out the statements for the boundary value problem. We define a compact simply connected domain $G \subset R^3$ with a border $S$. We accept that

$$\left| \partial_{\vec{r}}^{\alpha} \vec{v}_0(\vec{r}) \right| \leq C_{\alpha,K} \cdot (1 + |\vec{r}|)^{-K} \text{ on } G \subset R^3, \text{ for any } \alpha \text{ and } K, \tag{13}$$

$$\left| \partial_{\vec{r}}^{\alpha} \partial_t^m \vec{f}_{ex}(t,\vec{r}) \right| \leq C_{\alpha,m,K} \cdot (1 + |\vec{r}| + t)^{-K} \text{ on } [0,T] \times G \subset R^3 \ (T < \infty), \text{ for any } \alpha, m \text{ and } K, \tag{14}$$

$$\vec{v}(t,\vec{r}), \ p(t,\vec{r}), \ \vec{f}_{ex}(t,\vec{r}) \in C^{\infty}([0,T] \times G \subset R^3), \ \vec{v}_0(\vec{r}) \in C^{\infty} \ (\vec{r} \in G \subset R^3), \tag{15}$$

$$\int_G \left| \vec{v}(t,\vec{r}) \right|^2 \cdot d\vec{r} < const \text{ for } 0 \leq t \leq T \text{ (bounded energy).} \tag{16}$$

The initial step of the solution existence proof is the definition of a class of generalized solutions for which a uniqueness theorem is performed (Ladyzhenskaya, 2003).

Let's designate all functions satisfying the condition (16) as $L_2(G)$. As is known from the differential equation theory (Vladimirov, 1983), any function from this space can be presented as a Fourier series converging to this function. Due to the condition (15), the Fourier decompositions to converging series can be applied both to space and time variables. So we introduce the function space $L_2(G_T)$, $G_T = [0,T] \times G \subset [0,\infty) \times R^3$ (Ladyzhenskaya (2003), whose functions are decomposed to the Fourier series that are constructed from the complete system of functions $\{\vec{u}_n(t,\vec{r})\}$, $n = 0, \infty$. If this complete system is a Navier-Stokes equation solution the functions $\vec{v}(t,\vec{r}), p(t,\vec{r}) \in L_2(G_T), G_T = [0,T] \times G \subset [0,\infty) \times R^3$ presented as the expansions to this complete system are solutions of these equations.

A Fourier series is based on the use of the trigonometric functions which are presented as the exponential functions with the imaginary-valued argument. So a Fourier series can be presented as a complex-valued power series (Fihtengoltz, 2003a). For a continuously differentiable function on $[0,\infty) \times R^3$ the Lipschitz condition is performed for any domain $G_T'$, which is wider than $G_T$ ($G_T \subset G_T' \subset [0,\infty) \times R_3$), so the Fourier series of this function is uniformly converged on $G_T$ (Fihtengoltz, 2003a). The Fourier coefficients are uniquely determined from the boundary condition. So the local existence and uniqueness theorem can be applied to a general partial differential equation with constant coefficients (Courant & Hilbert, 1962) for the function space $G_T$.

We have the following reasons to suppose that this claim is also valid for the nonlinear Navier-Stokes equations. According to conditions (13)-(16), the functions $\vec{v}(t,\vec{r})$, $p(t,\vec{r})$, $\vec{v}_0(\vec{r})$ and $\vec{f}_{ex}(t,\vec{r})$ are $k$ times differentiable for $k=1,\dots,\infty$, and continuous with all their $k$th-order partial derivatives on $[0,\infty)\times R^3$. The nonlinear term of the Cauchy momentum equation is $k$ times differentiable and continuous together with its all derivatives also. Hence, $\vec{v}(t,\vec{r})$, $p(t,\vec{r})$, $\vec{v}_0(\vec{r})$, $\vec{f}_{ex}(t,\vec{r})$, their $k$th derivatives, and the nonlinear term can be presented as the uniformly converging Fourier series on $G_T$ (Fihtengoltz, 2003a). These Fourier series are differentiable and can be multiplied to other Fourier series. The result of multiplication is a Fourier series (Fihtengoltz, 2003a). The Fourier coefficients of the nonlinear term are uniquely defined from the boundary condition. Assuming that the nonlinear term contains no unknowns, we obtain the unique solution of the Navier-Stokes equations as for the system of two linear equations, one of which contains the additional source presented as the Fourier series.

We can obtain the unique global solution of the initial value problem (1), (2), and (3) satisfying the statement (A) of the Clay Mathematics Institute if we take the limit of the Fourier series for $G_T = [0,T] \times G \to [0,\infty) \times R^3$. As a result, the Fourier series turns to the Fourier transforms which are presented by improper integrals. Under conditions (5)-(8) these integrals exist and converge. The similar statements can be carried out for the periodic functions satisfying (9)-(12).

The use of a trigonometric Fourier series is not the only option to obtain a solution of the Stokes-Navier equations. It is possible that for some variables (e.g. time variable) the solution is decomposed to function series based on exponential functions with a real-valued argument instead of a complex-valued one. These series are real-valued power one. Under the condition of the uniform convergence we can reduce the nonlinear problem to a linear one and transform the series to improper integrals the same way as using the trigonometric Fourier series.

Let's verify whether the functions expandable to the uniformly converging Fourier series are really solutions of the Navier-Stokes equations. We will obtain the eigensolutions of the Navier-Stokes equations and study their relation to the Fourier series.

If $p(t,\vec{r}) \in L_2(G_T)$, for the proof of Navier-Stokes equation resolvability, it is enough to find the solutions of the mass continuity equation in $L_2(G_T)$. This is a constant coefficient homogeneous differential equation with private derivatives of the first order. It is linear and only one required unknown function $\vec{v}(t,\vec{r})$ is presented in it.

The Cauchy momentum equation is also linear with respect to the function $p(t,\vec{r})$. If we take the divergence of the right and left parts of Equation (1), we obtain that the pressure is a solution of the Poisson equation

$$\Delta p(t,\vec{r}) = div\vec{f}_{ex}(t,\vec{r}) - div(\rho \cdot (\vec{v}(t,\vec{r}) \cdot \nabla) \cdot \vec{v}(t,\vec{r})) \,. \tag{17}$$

The solution $p(t,\vec{r}) \in L_2(G_T)$ can be easily obtained at substitution of $\vec{v}(t,\vec{r})$ in this equation.

We should take into account that, generally, the velocity vector does not coincide with a direction of pressure gradient in the mass continuity equation. Therefore, we are interested in the resolvability of the Cauchy momentum equation in the absence of the pressure gradient. Thus, it is necessary to solve the equation

$$\rho \cdot \partial \vec{v}_\perp(t,\vec{r})/\partial t + \rho \cdot (\vec{v}(t,\vec{r}) \cdot \nabla) \cdot \vec{v}_\perp(t,\vec{r}) = \eta \cdot \Delta \vec{v}_\perp(t,\vec{r}) + \vec{f}_{ex,\perp}(t,\vec{r}), \tag{18}$$

where $\vec{v}(t,\vec{r}) = \vec{v}_{//}(t,\vec{r}) + \vec{v}_\perp(t,\vec{r})$, $\vec{f}_{ex}(t,\vec{r}) = \vec{f}_{ex,//}(t,\vec{r}) + \vec{f}_{ex,\perp}(t,\vec{r})$, $\vec{v}_{//}(t,\vec{r}) \| \nabla p(t,\vec{r}), \vec{v}_\perp(t,\vec{r}) \perp \nabla p(t,\vec{r})$, $\vec{f}_{ex,//}(t,\vec{r}) \| \nabla p(t,\vec{r})$, $\vec{f}_{ex,\perp}(t,\vec{r}) \perp \nabla p(t,\vec{r})$ under the conditions (2), (3), and (5)-(8).

## 2.3. Solutions of mass continuity equation

Let's consider the mass continuity equation $div(\vec{v}(t,\vec{r})) = 0$. This equation defines a velocity vector field as a divergence-free one.

For a divergence-free field,

$$curl(\vec{v}_{SOL}(t,\vec{r})) \neq 0 \, , \vec{v}_{SOL}(t,\vec{r})) = curl(\vec{H}(t,\vec{r})) \qquad (19)$$

where $\vec{H}(t,\vec{r})$ is a vector potential of a vector velocity field.

Let's admit that

$$curl(\vec{v}_{SOL}(t,\vec{r})) = \vec{j}(t,\vec{r}) \, , \; \vec{r} \in G \subset R^3 \qquad (20)$$

where $\vec{j}(t,\vec{r})$ is an arbitrary function under the condition $\vec{j}(t,\vec{r}) \in L_2(G), G \subset R^3$. In particular, $G \equiv R^3$. For this function,

$$div(\vec{j}(t,\vec{r})) = 0 \, . \qquad (21)$$

This expression is a necessary condition of resolvability of Equation (20) (Fihtengoltz, 2003a).

Let's obtain the general solution of the heterogeneous system of the equations:

$$curl(\vec{v}_{SOL}(t,\vec{r})) = \vec{j}(t,\vec{r}) \, , \; div_{SOL}(\vec{v}(t,\vec{r})) = 0 \, . \qquad (22)$$

We use the solution of the inverse problem of the vector analysis (Fihtengoltz, 2003a).

At first, we obtain the general solution of homogeneous system of the equations:

$$div(\vec{v}_{SOL}(t,\vec{r})) = 0 \, , \; curl(\vec{v}_{SOL}(t,\vec{r})) = 0 \, . \qquad (23)$$

The first equation testifies that a divergence-free field is a potential one:

$$\vec{v}_{SOL}(t,\vec{r})) = \vec{v}_{POT}(t,\vec{r})) = \nabla \Phi(t,\vec{r}) \qquad (24)$$

where $\Phi(t,\vec{r})$ is a scalar potential of a vector velocity field.

As a result, we obtain the Laplace equation

$$\Delta \Phi(t,\vec{r}) = 0 \, . \qquad (25)$$

Let's take a gradient of this expression. We consider that

$$\nabla \Delta \Phi(t,\vec{r}) = \Delta \nabla \Phi(t,\vec{r}) = \Delta \vec{v}_{POT}(t,\vec{r}) \, . \qquad (26)$$

We obtain the Laplace equation for the velocity:

$$\Delta \vec{v}_{POT}(t,\vec{r}) = 0 \, . \qquad (27)$$

The solutions of this equation are harmonic functions.

Let's obtain the private solution of the heterogeneous system (22). We substitute expression (19) for the vector potential in Equation (20).

Then,

$$curl(curl(\vec{H}(t,\vec{r})) = \vec{j}(t,\vec{r}) \Rightarrow$$
$$\nabla(div(\vec{H}(t,\vec{r})) - \Delta \vec{H}(t,\vec{r}) = \vec{j}(t,\vec{r}) \qquad (28)$$

where $\vec{H}(t,\vec{r})$ is a vector potential of a vector velocity field.

Using the correlations $curl(\nabla(div(\vec{H}(t,\vec{r}))) = 0$ and $curl(\Delta \vec{H}(t,\vec{r})) = \Delta(curl \vec{H}(t,\vec{r})) = \Delta \vec{v}_{SOL}(t,\vec{r})$, we obtain the Poisson equation for the velocity

$$\Delta \vec{v}_{SOL}(t,\vec{r}) = -\nabla \times \vec{j}(t,\vec{r}) \, . \qquad (29)$$

The Poisson equation for the vector potential is

$$\Delta \vec{H}(t,\vec{r}) = -\vec{j}(t,\vec{r}) \, , \; div(\vec{H}(t,\vec{r})) = 0 \, . \qquad (30)$$

The general solution of the heterogeneous Poisson equation (29) includes the solution of the Laplace equation (27) as the general solution of the homogeneous equation. Thus, the general solution of the system (22) is reduced to the general solution of Equation (29).

The vector field $\vec{v}(t,\vec{r})$ is completely defined by its scalar and vector potentials. Let's investigate the ability of potentials $\Phi(t,\vec{r})$ and $H(t,\vec{r})$ to build the complete systems of functions in three-dimensional space $R^3$ if these potentials are harmonic functions.

The harmonic functions can be presented as converging power series in three dimensions. For two space variables these power series have the form of the trigonometric Fourier series. For the third variable, the power series is based on the exponential function with the real-valued argument.

The harmonic functions possess the property of completeness only for two variables. For the third variable it is impossible to construct the complete system of functions.

For example, let's consider solutions $\Phi(t,\vec{r})$ of the three-dimensional Laplace equation in the rectangular Cartesian coordinates. The solutions in the form of linear functions aren't interesting to us. We investigate the solution which is looking as $\Phi(t,\vec{r}) \sim \exp(\vec{k} \cdot \vec{r})$ for $\vec{k}^2 = 0$, where $\vec{k}$ is a complex-valued vector. In this solution the value of one wave vector projection always depends on two others. The correlation between projections does not allow constructing the orthonormal basis for one of the three spatial coordinates. The complete system can be constructed only for the functions of two spatial variables.

In the cylindrical coordinates the solutions of the Laplace equation are functions like

$$\Phi(t,\vec{r}) \sim R_n(k_\rho \cdot \rho) \cdot \exp(\pm ik_\varphi \cdot \varphi) \cdot \exp(-k_\rho \cdot z) \tag{31}$$

where $R_n(k_\rho \cdot \rho)$ is the cylindrical Bessel function and $k_\rho$ and $k_\varphi$ are the real eigenvalues of the problem. As well as in the previous case, we have only two independent eigenvalues for the function of three variables.

In the spherical coordinates the solution aspiring zero at infinity looks like

$$\Phi(t,\vec{r}) \sim (1/r^n) \cdot Y_n(\vartheta, \varphi) \tag{32}$$

where $Y_n(\vartheta, \varphi)$ is the spherical superficial harmonics possessing the property of completeness.

The functions $1/r^n$ don't possess this property. The correlation between the eigenvalues in the Cartesian coordinate problem is kept also in the curvilinear one. The solutions depending on the curvilinear coordinates can be obtained by the expansion of the exponential function to the series on the complete system of the functions corresponding defined geometry (e.g. the cylindrical and spherical functions). Under the condition of $\vec{k}^2 = 0$ these functions lose the completeness for three variables.

But it is much more important that the harmonic functions represent a nonviscous fluid motion. According to formula (27), there is no a term with the viscosity in the Cauchy momentum equation. A viscous fluid motion is defined by not a potential (curl-free) divergence-free field but exclusively a divergence-free one which is determined by heterogeneous Equation (30).

Let's consider the case when the velocity $\vec{v}(t,\vec{r})$ is defined only by the vector potential $\vec{H}(t,\vec{r})$. We substitute the functions

$$\vec{v}(t,\vec{r}) = \vec{v}(t,\vec{k}) \cdot \exp(\vec{k} \cdot \vec{r}), \tag{33}$$

$$\vec{j}(t,\vec{r}) = \vec{j}(t,\vec{k}) \cdot \exp(\vec{k} \cdot \vec{r}) \tag{34}$$

in Equation (29). We obtain

$$\vec{v}(t,\vec{k}) = \frac{\vec{j}(t,\vec{k}) \times \vec{k}}{\vec{k}^2}, \ \vec{k}^2 \neq 0. \tag{35}$$

It means that

$$\vec{k} \perp \vec{v}(t,\vec{r}). \tag{36}$$

We have the same result if we substitute the function (33) in the equation $div(\vec{v}(t,\vec{r})) = 0$. We obtain the expression

$$\vec{k} \cdot \vec{v}(t,\vec{r}) = 0. \tag{37}$$

The case when

$$\vec{v}(t,\vec{r}) \sim v(t,\vec{k}) \cdot (\vec{k}/k) \tag{38}$$

leads to the expression $\vec{k}^2 = 0$. It is performed for the potential velocity $\vec{v}_{POT}(t,\vec{r})$ satisfying Equation (27).

The alternative solution is $\vec{k} \perp \vec{v}(t,\vec{r})$ where $\vec{k}$ is a complex-valued vector. It is valid for the velocity $\vec{v}_{SOL}(t,\vec{r})$ satisfying Equation (29). In this case, there exists correlation defined by the necessity to choose the vector $\vec{k}$ direction that is perpendicular to the vector $\vec{v}(t,\vec{r})$. For the problem solving in the Cartesian or curvilinear coordinates, this correlation reduces the number of uncertain eigenvalues as it has occurred for the Laplace equation. If one axis of the Cartesian system of coordinates is directed in parallel to the vector $\vec{v}(t,\vec{r})$ the solution becomes the function of only two spatial variables. The transformation to an arbitrary Cartesian system of coordinates and a curvilinear coordinate system leads to the result that there are only two independent spatial variables.

## 2.4. Solutions of Cauchy momentum equation

Under the condition $\eta > 0$ the three-dimensional Navier-Stokes equations are reduced to a two-dimensional one. Besides this, the nonlinear term is equal to 0 in the Cauchy momentum equation if there are no external sources of a fluid motion in $R^3$. Using expression (33), we have

$$(\vec{v}(t,\vec{r}) \cdot \nabla) \cdot \vec{v}(t,\vec{r}) = (\vec{v}_{SOL}(t,\vec{r}) \cdot \nabla) \cdot \vec{v}_{SOL}(t,\vec{r}) \sim (\vec{k}' \cdot \vec{v}_{SOL}(t,\vec{k})) \cdot \vec{v}_{SOL}(t,\vec{k}') = 0 \qquad (39)$$

because the condition $\vec{k} \perp \vec{v}(t,\vec{r})$ is performed for any $\vec{k}$ in $\vec{v}_{SOL}(t,\vec{k})$.

It means that there is no a term with velocity in Equation (17):

$$\Delta p(t,\vec{r}) = div \vec{f}_{ex}(t,\vec{r}), \qquad (40)$$

and Equation (18) is a linear one:

$$\rho \cdot \partial \vec{v}_{SOL}(t,\vec{r})/\partial t = \eta \cdot \Delta \vec{v}_{SOL}(t,\vec{r}) + \vec{f}_{ex}(t,\vec{r}). \qquad (41)$$

As a result, we obtain the eigensolution of the Navier-Stokes equations for velocity in the Cartesian coordinate system with unit vectors $\vec{e}_x$, $\vec{e}_y$, and $\vec{e}_z$ under the condition $\eta > 0$ as

$$\vec{v}_{SOL}(t,\vec{r}) = \vec{v}_{SOL}(\vec{k}) \cdot \exp[-\eta \cdot \frac{\vec{k}^2}{\rho}] \cdot \exp(i \cdot \vec{k}\vec{r}) =$$

$$\exp[-\eta \cdot \frac{(k'_y)^2 + (k'_z)^2}{\rho} \cdot t] \cdot v_x(k'_y, k'_z) \cdot \exp[i \cdot (k'_y \cdot y + k'_z \cdot z)] \cdot \vec{e}_x +$$

$$\exp[-\eta \cdot \frac{(k''_x)^2 + (k''_z)^2}{\rho} \cdot t] \cdot v_y(k''_x, k''_z) \cdot \exp[i \cdot (k''_x \cdot x + k''_z \cdot z)] \cdot \vec{e}_y + \qquad (42)$$

$$\exp[-\eta \cdot \frac{(k'''_x)^2 + (k'''_y)^2}{\rho} \cdot t] \cdot v_z(k'''_x, k'''_y) \cdot \exp[i \cdot (k'''_x \cdot x + k'''_y \cdot y)] \cdot \vec{e}_z$$

where $\vec{k} = (k_x, k_y, k_z)$ and $\vec{r} = (x, y, z)$ are the real-valued vectors, $\vec{k} \perp \vec{v}(t,\vec{r})$. This formula is obtained using the expressions:

$$k_x = f_1(k_y, k_z), k_y = f_2(k_x, k_z), \ k_z = f_3(k_x, k_y), \ x = f_1(y, z), y = f_2(x, z), \ z = f_3(x, y) \qquad (43)$$

where $f_1$, $f_2$, and $f_3$ are linear functions.

We obtain the global solution of the problem for $\vec{f}_{ex}(t,\vec{r}) \equiv 0$ if we take the improper integral of the function (42) over the continuous spectra of $k'_y, k'_z, k''_x, k''_z, k'''_x, k'''_y$ and satisfy the initial condition. The solution is a smooth function at any time moment ($0 \le t < \infty$).

For pressure, we have the general solution of the two-dimensional Poisson equation (40).

We should say that the nonlinear term isn't equal to 0 if we use the curvilinear coordinates (e.g. spherical or cylindrical) instead of the Cartesian one. In this case, we implicitly use a source of a fluid motion at some points of space $R^3$ (e.g. a point or line source). But $div(\vec{v}(t,\vec{r})) \neq 0$ at the source points.

## 3. Alternative

The divergence-free velocity field defines viscous two-dimensional solutions of the Navier-Stokes equations. It is necessary to update the Navier-Stokes equations to consider a potential field which is not divergence-free for solving a viscous three-dimensional problem.

The necessity of updating the equations is most evident at the consideration of fluid movement near body surfaces. Sticking fluid molecules to a surface means inelastic collisions of these molecules with a body. The molecules colliding with the stuck one are also involved in the inelastic collisions since they stick to other molecules or lose kinetic energy equal to the binding energy to release the attached molecules. Therefore there are inelastic collisions in the whole boundary layer.

The mass continuity equation does not include the inelastic processes. It means that the field of velocities is divergence-free. It is natural to expect that the divergence-free field does not possess sufficient capabilities to describe the fluid motion in the boundary layer. We suppose that the inelastic intermolecular interactions can be taken into account in the equations by the use of the potential field where velocity divergence is not equal to zero. It distinguishes it from the velocity potential field which is already used in fluid dynamics. That field is a kind of a divergence-free one.

## 3.1. Modification of equations

Let's update the Navier-Stokes equations using a curl-free potential vector field. The refined mass continuity equation looks as follows

$$div(\vec{v}(t,\vec{r})) = \Delta \Phi(t,\vec{r}), \tag{44}$$

$$\vec{v}(t,\vec{r}) = \vec{v}_{SOL}(t,\vec{r}) + \vec{v}_{POT}(t,\vec{r}), \; \vec{v}_{POT}(t,\vec{r}) = \nabla \Phi(t,\vec{r}), \vec{v}_{SOL}(t,\vec{r})) = curl(\vec{H}(t,\vec{r})) \tag{45}$$

where $\Phi(t,\vec{r})$ is the scalar potential and $\vec{H}(t,\vec{r})$ is the vector potential of the vector velocity field.

Equation (44) always has solutions. Generally, their number is infinite (Fihtengoltz, 2003a).

We use the general form of the Cauchy momentum equation

$$\rho \cdot \partial \vec{v}(t,\vec{r}) / \partial t + \rho \cdot (\vec{v}(t,\vec{r}) \cdot \nabla) \cdot \vec{v}(t,\vec{r}) = -\nabla p(t,\vec{r}) + \eta \cdot \Delta \vec{v}(t,\vec{r}) + (\eta/3 + \zeta) \cdot \nabla(\nabla \vec{v}(t,\vec{r})) + \vec{f}_{ex}(t,\vec{r}) \tag{46}$$

where $\zeta$ is the second viscosity considering the intermolecular interactions. The compressibility of a fluid is not used in the derivation of this equation (Landau & Lifshitz, 1987), i.e. we can accept that $\rho = const$.

We substitute expression (44) in the term $(\eta/3 + \zeta) \cdot \nabla(\nabla \vec{v}(t,\vec{r}))$ of the Cauchy momentum equation (46). We take into account that

$$\nabla \Delta \Phi(t,\vec{r}) = \Delta \nabla \Phi(t,\vec{r}) = \Delta \vec{v}_{POT}(t,\vec{r}). \tag{47}$$

We obtain the Cauchy momentum equation in the form

$$\rho \cdot \partial \vec{v}(t,\vec{r}) / \partial t + \rho \cdot (\vec{v}(t,\vec{r}) \cdot \nabla) \cdot \vec{v}(t,\vec{r}) = -\nabla p(t,\vec{r}) + \eta \cdot \Delta \vec{v}_{SOL}(t,\vec{r}) + (4\eta/3 + \zeta) \cdot \Delta \vec{v}_{POT}(t,\vec{r})) + \vec{f}_{ex}(t,\vec{r}). \tag{48}$$

Any vector field can be presented as a sum of a divergence-free and a curl-free vector field (Fihtengoltz, 2003a). We suppose that

$$\vec{f}_{ex}(t,\vec{r}) = \vec{f}_{ex,POT}(t,\vec{r}) + \vec{f}_{ex,SOL}(t,\vec{r}), \; \vec{f}_{ex,POT}(t,\vec{r}) = \nabla \Phi_{ex}(t,\vec{r}), \; \vec{f}_{ex,SOL}(t,\vec{r}) = curl(\vec{H}_{ex}(t,\vec{r})) \tag{49}$$

where $\Phi_{ex}(t,\vec{r})$ is the scalar potential and $\vec{H}_{ex}(t,\vec{r})$ is the vector potential of the external source vector field.

The vector field $\nabla p(t,\vec{r})$ is potential. So expression (48) can be separated into two equations

$$\rho \cdot \partial \vec{v}_{POT}(t,\vec{r}) / \partial t + \rho \cdot (\vec{v}(t,\vec{r}) \cdot \nabla) \cdot \vec{v}_{POT}(t,\vec{r}) = -\nabla p(t,\vec{r}) + (4\eta/3 + \zeta) \cdot \Delta \vec{v}_{POT}(t,\vec{r})) + \vec{f}_{ex,POT}(t,\vec{r}), \tag{50}$$

$$\rho \cdot \partial \vec{v}_{SOL}(t,\vec{r}) / \partial t + \rho \cdot (\vec{v}(t,\vec{r}) \cdot \nabla) \cdot \vec{v}_{SOL}(t,\vec{r}) = \eta \cdot \Delta \vec{v}_{SOL}(t,\vec{r}) + \vec{f}_{ex,SOL}(t,\vec{r}). \tag{51}$$

We consider Equation (50). We use the expressions (Landau & Lifshitz, 1987)

$$\vec{v}_{POT}(t,\vec{r}) = \nabla \Phi(t,\vec{r}),$$

$$(\vec{v}_{POT}(t,\vec{r}) \cdot \nabla) \cdot \vec{v}_{POT}(t,\vec{r}) = \frac{1}{2} \cdot \nabla \vec{v}_{POT}^2(t,\vec{r}) - \vec{v}_{POT}(t,\vec{r}) \times curl(\vec{v}_{POT}(t,\vec{r})) = \frac{1}{2} \cdot \nabla \vec{v}_{POT}^2(t,\vec{r}).$$

We obtain the solution for pressure

$$p(t,\vec{r}) = \tilde{p}(t,\vec{r}) + \rho \cdot [-\partial \Phi(t,\vec{r})/\partial t - \frac{(\nabla \Phi(t,\vec{r}))^2}{2} + \frac{1}{\rho} \cdot (\frac{4 \cdot \eta}{3} + \zeta) \cdot \Delta \Phi(t,\vec{r})] + f(t) + \Phi_{ex}(t,\vec{r}) \quad (52)$$

where $f(t)$ is an arbitrary function of time and $\tilde{p}(t,\vec{r})$ is the solution of the equation

$$\Delta \tilde{p}(t,\vec{r}) = -\rho \cdot div[(\vec{v}_{SOL}(t,\vec{r}) \cdot \nabla) \cdot \vec{v}_{POT}(t,\vec{r})] \,. \quad (53)$$

In presence of the divergence-free and curl-free fields, the system of the modified Navier-Stokes equations is

$$\vec{v}(t,\vec{r}) = \vec{v}_{SOL}(t,\vec{r}) + \vec{v}_{POT}(t,\vec{r}), \;\; \vec{f}_{ex}(t,\vec{r}) = \nabla \Phi_{ex}(t,\vec{r}) + \vec{f}_{ex,SOL}(t,\vec{r})$$

$$p(t,\vec{r}) = \tilde{p}(t,\vec{r}) + \rho \cdot [-\partial \Phi(t,\vec{r})/\partial t - \frac{(\nabla \Phi(t,\vec{r}))^2}{2} + \frac{1}{\rho} \cdot (\frac{4 \cdot \eta}{3} + \zeta) \cdot \Delta \Phi(t,\vec{r})] + f(t) + \Phi_{ex}(t,\vec{r}), \quad (54)$$

$$\Delta \tilde{p}(t,\vec{r}) = -\rho \cdot div[(\vec{v}_{SOL}(t,\vec{r}) \cdot \nabla) \cdot \vec{v}_{POT}(t,\vec{r})], \quad (55)$$

$$\rho \cdot \partial \vec{v}_{SOL}(t,\vec{r})/\partial t + \rho \cdot (\vec{v}(t,\vec{r}) \cdot \nabla) \cdot \vec{v}_{SOL}(t,\vec{r}) = \eta \cdot \Delta \vec{v}_{SOL}(t,\vec{r}) + \vec{f}_{ex,SOL}(t,\vec{r}), \quad (56)$$

$$div(\vec{v}(t,\vec{r})) = \Delta \Phi(t,\vec{r}) \,. \quad (57)$$

These equations should be complemented by conditions (3) and (4).

The existence of the general solution of Equation (56) is proved in the next part. The whole system (54-57) with conditions (3) and (4) is solvable in three dimensions. The Fourier method allows obtaining the smooth solutions of these equations.

## 3.2. Incompressibility of fluid

If we accept that [×] a crucial problem is whether the fluid is incompressible. According to the work of Batchelor (2000), the influence of pressure variations on the value of fluid density is negligible if

$$\left| \frac{1}{\rho} \cdot \frac{D\rho}{Dt} \right| << \frac{U}{L}, \text{ or } \left| div(\vec{v}(t,\vec{r})) \right| << \frac{U}{L} \quad (58)$$

where $L$ is the length scale ($\vec{v}(t,\vec{r})$ varies slightly over distances small compared with the scale), $U$ is the value of the variations of $\left| \vec{v}(t,\vec{r}) \right|$ with respect to both position and time. *This means that the velocity of an incompressible fluid is only approximately divergence-free.*

In our research we suppose that assumption (58) is valid, i.e. a fluid is incompressible. We only accept the existence of a curl-free velocity field besides a divergence-free one on the condition that

$$div(\vec{v}_{SOL}(t,\vec{r})) = 0, \; \left| div(\vec{v}_{POT}(t,\vec{r})) \right| << \frac{U}{L}, \text{ but } div(\vec{v}_{POT}(t,\vec{r})) \neq 0 \,. \quad (59)$$

It is performed for a nonstationary flow if (Landau & Lifshitz, 1987)

$$\left| \vec{v}(t,\vec{r}) \right| << c_s, \quad (60)$$

$$\tau >> \frac{L}{c_s} \quad (61)$$

where $c_s$ is the speed of sound in the fluid and $\tau$ is the time scale. The last condition means that the time during which the sound passes the distance equal to the length scale has to be much less that of the time scale $\tau$ during which the motion of the fluid significantly changes, i.e. propagation of interactions in the fluid is instantaneous.

The condition (59) is important if we want to investigate the nonlinear phenomena of an incompressible fluid motion. Small terms in nonlinear equations can have a significant influence on behavior of a physical system in space and time as a result of a cumulative effect (Bogoliubov & Mitropolski, 1961). The equations of the viscous fluid motion become more accurate at the account of curl-free vector field.

## 3.3. Existence of solutions

Let's solve Equation (56):

$$\partial \vec{v}_{SOL}(t,\vec{r})/\partial t + (\vec{v}(t,\vec{r}) \cdot \nabla) \vec{v}_{SOL}(t,\vec{r}) = \eta \cdot \Delta \vec{v}_{SOL}(t,\vec{r}) + \vec{f}_{ex,SOL}(t,\vec{r}),$$ (62)

At first let's consider this equation without $\vec{f}_{ex,SOL}(t,\vec{r})$:

$$\partial \vec{v}_{SOL}(t,\vec{r})/\partial t + (\vec{v}(t,\vec{r}) \cdot \nabla) \vec{v}_{SOL}(t,\vec{r}) = \eta \cdot \Delta \vec{v}_{SOL}(t,\vec{r}).$$ (63)

We define $\vec{v}_{SOL}(t,\vec{r})$ and $\vec{v}(t,\vec{r})$ as a uniformly converging series

$$\vec{v}_{SOL}(t,\vec{r}) \equiv \vec{u}(t,\vec{r}) = \sum_{j=1}^{\infty} \vec{u}_j(t,\vec{r}), \quad \vec{v}(t,\vec{r}) = \sum_{j=1}^{\infty} \vec{v}_j(t,\vec{r}),$$ (64)

$$\upsilon = \frac{\eta}{\rho}.$$ (65)

$$\partial \vec{u}_1(t,\vec{r})/\partial t = \upsilon \cdot \Delta \vec{u}_1(t,\vec{r}),$$
$$\partial \vec{u}_2(t,\vec{r})/\partial t + (\vec{v}_1(t,\vec{r}) \cdot \nabla) \cdot \vec{u}_1(t,\vec{r}) = \upsilon \cdot \Delta \vec{u}_2(t,\vec{r}),$$
$$\partial \vec{u}_3(t,\vec{r})/\partial t + (\vec{v}_2(t,\vec{r}) \cdot \nabla) \cdot \vec{u}_1(t,\vec{r}) + (\vec{v}_1(t,\vec{r}) \cdot \nabla) \cdot \vec{u}_2(t,\vec{r}) = \upsilon \cdot \Delta \vec{u}_3(t,\vec{r}),$$ (66)
...

$$\partial \vec{u}_j(t,\vec{r})/\partial t + \sum_{m=1}^{j-1} (\vec{v}_{j-m}(t,\vec{r}) \cdot \nabla) \cdot \vec{u}_m(t,\vec{r}) = \upsilon \cdot \Delta \vec{u}_j(t,\vec{r}),$$

....

We obtained the infinite set of the inhomogeneous linear heat equations. The existence and uniqueness theorem is proved for them (Tikhonov, 2013). The infinite series of private solutions for infinite number of the heat equations converges uniformly on $[0,\infty) \times R^3$. So we can obtain the unique solution of Equation (56) if the initial and boundary conditions are defined.

The presence $\vec{f}_{ex,SOL}(t,\vec{r})$ in the right part of the equation doesn't change the reasoning. But we will show that there exists a turbulent solution of Equation (56) besides the unique solution for a laminar flow.

At the beginning we obtain the solution of Equation (56) for a laminar flow which can blow up as a result of a resonance effect.

### 3.4. Resonant blowup of laminar flow

We consider the equation

$$\partial \vec{u}(t,\vec{r})/\partial t + (\vec{v}(t,\vec{r}) \cdot \nabla) \vec{u}(t,\vec{r}) = \upsilon \cdot \Delta \vec{u}(t,\vec{r}).$$ (67)

Let's suppose that

$$\vec{v}(t,\vec{r}) \equiv \vec{v}_1(t,\vec{r}).$$ (68)

to simplify the expressions.

We obtain the system

$$\partial \vec{u}_1(t,\vec{r})/\partial t = \upsilon \cdot \Delta \vec{u}_1(t,\vec{r}),$$
$$\partial \vec{u}_2(t,\vec{r})/\partial t + (\vec{v}(t,\vec{r}) \cdot \nabla) \cdot \vec{u}_1(t,\vec{r}) = \upsilon \cdot \Delta \vec{u}_2(t,\vec{r}),$$
$$\partial \vec{u}_3(t,\vec{r})/\partial t + (\vec{v}(t,\vec{r}) \cdot \nabla) \cdot \vec{u}_2(t,\vec{r}) = \upsilon \cdot \Delta \vec{u}_3(t,\vec{r}),$$ (69)
...

$$\partial \vec{u}_j(t,\vec{r})/\partial t + (\vec{v}(t,\vec{r}) \cdot \nabla) \cdot \vec{u}_{j-1}(t,\vec{r}) = \upsilon \cdot \Delta \vec{u}_j(t,\vec{r}),$$

....

We define the solutions in the form

$$v(t,\vec{r}) = \exp(-\omega_0^2 \upsilon \cdot t) \cdot \vec{v}(\vec{r}),$$ (70)

$$\vec{u}_j(t,\vec{r}) = \exp(-[(j-1) \cdot \omega_0^2 + \lambda_0^2] \cdot \upsilon \cdot t) \cdot \vec{u}_j(\vec{r}), \quad j = 1,2,\ldots,\infty.$$ (71)

We obtain the linear system of the equations

$$\Delta \vec{u}_1(\vec{r}) + \lambda_0^2 \cdot \vec{u}_1(\vec{r}) = 0,$$

$$\Delta \vec{u}_2(\vec{r}) + (\omega_0^2 + \lambda_0^2) \cdot \vec{u}_2(\vec{r}) = \frac{(\vec{v}(\vec{r}) \cdot \nabla) \cdot \vec{u}_1(\vec{r})}{\upsilon},$$

$$\Delta \vec{u}_3(\vec{r}) + (2\omega_0^2 + \lambda_0^2) \cdot \vec{u}_3(\vec{r}) = \frac{(\vec{v}(\vec{r}) \cdot \nabla) \cdot \vec{u}_2(\vec{r})}{\upsilon}, \qquad (72)$$

$$\dots$$

$$\Delta \vec{u}_j(\vec{r}) + [(j-1) \cdot \omega_0^2 + \lambda_0^2] \cdot \vec{u}_j(\vec{r}) = \frac{(\vec{v}(\vec{r}) \cdot \nabla) \cdot \vec{u}_{j-1}(\vec{r})}{\upsilon},$$

$$\dots.$$

The solution of the first equation is

$$\vec{u}_1(\vec{r}) = \vec{u}_0 \cdot \cos(\vec{k}\vec{r}), \ k^2 = \lambda_0^2. \qquad (73)$$

Let's obtain the solutions of the second and the third equation. We suppose that

$$\vec{v}(\vec{r}) = \vec{v}_0 \cdot \cos(\vec{k}_0 \vec{r}). \qquad (74)$$

$$\Delta \vec{u}_2(\vec{r}) + (\omega_0^2 + \lambda_0^2) \cdot \vec{u}_2(\vec{r}) = -\frac{(\vec{v}_0 \cdot \vec{k})}{\upsilon} \cdot \cos(\vec{k}_0 \vec{r}) \cdot \sin(\vec{k}\vec{r}) \cdot \vec{u}_0 = \qquad (75)$$

$$= -\frac{(\vec{v}_0 \cdot \vec{k})}{2 \cdot \upsilon} \cdot \{\sin[(\vec{k} + \vec{k}_0) \cdot \vec{r}] + \sin[(\vec{k} - \vec{k}_0) \cdot \vec{r}]\} \cdot \vec{u}_0.$$

$$\vec{u}_2(\vec{r}) = \frac{(\vec{v}_0 \cdot \vec{k}) \cdot \vec{u}_0}{2 \cdot \upsilon \cdot [(\vec{k} + \vec{k}_0)^2 - (\omega_0^2 + \lambda_0^2)]} \cdot \sin[(\vec{k} + \vec{k}_0) \cdot \vec{r}] + \frac{(\vec{v}_0 \cdot \vec{k}) \cdot \vec{u}_0}{2 \cdot \upsilon \cdot [(\vec{k} - \vec{k}_0)^2 - (\omega_0^2 + \lambda_0^2)]} \sin[(\vec{k} - \vec{k}_0) \cdot \vec{r}]. \quad (76)$$

$$\Delta \vec{u}_3(\vec{r}) + (2\omega_0^2 + \lambda_0^2) \cdot \vec{u}_3(\vec{r}) = \frac{[(\vec{v}_0 \cdot \vec{k})^2 + (\vec{v}_0 \cdot \vec{k}) \cdot (\vec{v}_0 \cdot \vec{k}_0)] \cdot \vec{u}_0}{4 \cdot \upsilon \cdot [(\vec{k} + \vec{k}_0)^2 - (\omega_0^2 + \lambda_0^2)]} \cdot \{\cos[(\vec{k} + 2\vec{k}_0) \cdot \vec{r}] + \cos(\vec{k}\vec{r})\} +$$

$$+ \frac{[(\vec{v}_0 \cdot \vec{k})^2 - (\vec{v}_0 \cdot \vec{k}) \cdot (\vec{v}_0 \cdot \vec{k}_0)] \cdot \vec{u}_0}{4 \cdot \upsilon \cdot [(\vec{k} - \vec{k}_0)^2 - (\omega_0^2 + \lambda_0^2)]} \cdot \{\cos(\vec{k}\vec{r}) + \cos[(\vec{k} - 2\vec{k}_0) \cdot \vec{r}]\}. \qquad (77)$$

$$\vec{u}_3(\vec{r}) = \frac{\vec{u}_0 \cdot \cos(\vec{k}\vec{r})}{4 \cdot \upsilon \cdot (2\omega_0^2 + \lambda_0^2 - \vec{k}^2)} \cdot \{\frac{(\vec{v}_0 \cdot \vec{k})^2 + (\vec{v}_0 \cdot \vec{k}) \cdot (\vec{v}_0 \cdot \vec{k}_0)}{(\vec{k} + \vec{k}_0)^2 - (\omega_0^2 + \lambda_0^2)} + \frac{(\vec{v}_0 \cdot \vec{k})^2 - (\vec{v}_0 \cdot \vec{k}) \cdot (\vec{v}_0 \cdot \vec{k}_0)}{(\vec{k} - \vec{k}_0)^2 - (\omega_0^2 + \lambda_0^2)}\} +$$

$$+ \frac{[(\vec{v}_0 \cdot \vec{k})^2 + (\vec{v}_0 \cdot \vec{k}) \cdot (\vec{v}_0 \cdot \vec{k}_0)] \cdot \vec{u}_0}{4 \cdot \upsilon \cdot (2\omega_0^2 + \lambda_0^2 - (\vec{k} + 2\vec{k}_0)^2) \cdot [(\vec{k} + \vec{k}_0)^2 - (\omega_0^2 + \lambda_0^2)]} \cdot \cos[(\vec{k} + 2\vec{k}_0) \cdot \vec{r}] + \qquad (78)$$

$$+ \frac{[(\vec{v}_0 \cdot \vec{k})^2 - (\vec{v}_0 \cdot \vec{k}) \cdot (\vec{v}_0 \cdot \vec{k}_0)] \cdot \vec{u}_0}{4 \cdot \upsilon \cdot (2\omega_0^2 + \lambda_0^2 - (\vec{k} - 2\vec{k}_0)^2) \cdot [(\vec{k} - \vec{k}_0)^2 - (\omega_0^2 + \lambda_0^2)]} \cdot \cos[(\vec{k} - 2\vec{k}_0) \cdot \vec{r}].$$

We study the nonlinear effect which is described as the interaction between the plane wave functions having the initial wave vectors $\vec{k}$ and $\vec{k}_0$. The third and higher order approximations of the solution obtain the term which wave vector coincident with the initial wave vector $\vec{k}$. It is the term which has the function $\cos(\vec{k}\vec{r})$. It means that there is a resonance which increases the amplitude of the solution with time (Landau & Lifshitz, 1976).

Let's study this effect more carefully.
We accept at the next step that

$$\vec{k}_0 = \vec{k}, \ \vec{k}^2 = \lambda_0^2, \ \vec{v}_0 \uparrow\uparrow \vec{k}. \qquad (79)$$

We obtain

$$\Delta \vec{u}_2(\vec{r}) + (\omega_0^2 + \lambda_0^2) \cdot \vec{u}_2(\vec{r}) = -\frac{v_0 \cdot k}{2 \cdot \upsilon} \cdot \sin(2\vec{k}\vec{r}) \cdot \vec{u}_0, \qquad (80)$$

$$\vec{u}_2(\vec{r}) = \frac{v_0 \cdot k \cdot \vec{u}_0}{2 \cdot \upsilon \cdot (3\lambda_0^2 - \omega_0^2)} \cdot \sin(2\vec{k}\vec{r}) \,. \tag{81}$$

$$\Delta \vec{u}_3(\vec{r}) + (2\omega_0^2 + \lambda_0^2) \cdot \vec{u}_3(\vec{r}) = \frac{v_0^2 \cdot \lambda_0^2 \cdot \vec{u}_0}{2 \cdot \upsilon^2 \cdot (3\lambda_0^2 - \omega_0^2)} \cdot [\cos(\vec{k}\vec{r}) + \cos(3\vec{k}\vec{r})], \tag{82}$$

$$\vec{u}_3(\vec{r}) = \frac{v_0^2 \cdot \lambda_0^2 \cdot \cos(\vec{k}\vec{r}) \cdot \vec{u}_0}{4 \cdot \upsilon^2 \cdot \omega_0^2 \cdot (3\lambda_0^2 - \omega_0^2)} + \frac{v_0^2 \cdot \lambda_0^2 \cdot \cos(3\vec{k}\vec{r}) \cdot \vec{u}_0}{4 \cdot \upsilon^2 \cdot (\omega_0^2 - 4\lambda_0^2) \cdot (3\lambda_0^2 - \omega_0^2)} \,. \tag{83}$$

Let's obtain the next order of solutions to establish the general form of the resonance amplitude. We are interested in the solutions only with $\cos(\vec{k}\vec{r})$. We define them as $\vec{u}_{2j+1}^1(\vec{r})$. We need also the solutions $\vec{u}_{2j}^1(\vec{r})$ with $\sin(2\vec{k}\vec{r})$ to obtain $\vec{u}_{2j+1}^1(\vec{r})$.

$$\vec{u}_3^1(\vec{r}) = \frac{v_0^2 \cdot \lambda_0^2 \cdot \cos(\vec{k}\vec{r}) \cdot \vec{u}_0}{2 \cdot \upsilon^2 \cdot 2\omega_0^2 \cdot (3\lambda_0^2 - \omega_0^2)} = \frac{v_0^2 \cdot \lambda_0^2 \cdot \cos(\vec{k}\vec{r}) \cdot \vec{u}_0}{2^2 \cdot \upsilon^2 \cdot \omega_0^2 \cdot (3\lambda_0^2 - \omega_0^2)} \,. \tag{84}$$

$$\Delta \vec{u}_4^1(\vec{r}) + (3\omega_0^2 + \lambda_0^2) \cdot \vec{u}_4^1(\vec{r}) = \frac{(\vec{v}(\vec{r}) \cdot \nabla) \cdot \vec{u}_3^1(\vec{r})}{\upsilon} \,, \tag{85}$$

$$\Delta \vec{u}_4^1(\vec{r}) + (3\omega_0^2 + \lambda_0^2) \cdot \vec{u}_4^1(\vec{r}) = -\frac{v_0^3 \cdot \lambda_0^3 \cdot \sin(2\vec{k}\vec{r}) \cdot \vec{u}_0}{2^2 \cdot \upsilon^3 \cdot 2\omega_0^2 \cdot (3\lambda_0^2 - \omega_0^2)} = -\frac{v_0^3 \cdot \lambda_0^3 \cdot \sin(2\vec{k}\vec{r}) \cdot \vec{u}_0}{2^3 \cdot \upsilon^3 \cdot \omega_0^2 \cdot (3\lambda_0^2 - \omega_0^2)} \,, \tag{86}$$

$$\vec{u}_4^1(\vec{r}) = \frac{v_0^3 \cdot \lambda_0^3 \cdot \sin(2\vec{k}\vec{r}) \cdot \vec{u}_0}{2^3 \cdot \upsilon^3 \cdot \omega_0^2 \cdot (3\lambda_0^2 - \omega_0^2) \cdot (3\lambda_0^2 - 3\omega_0^2)} \,. \tag{87}$$

$$\Delta \vec{u}_5^1(\vec{r}) + (4\omega_0^2 + \lambda_0^2) \cdot \vec{u}_5^1(\vec{r}) = \frac{(\vec{v}(\vec{r}) \cdot \nabla) \cdot \vec{u}_4^1(\vec{r})}{\upsilon} \,, \tag{88}$$

$$\Delta \vec{u}_5^1(\vec{r}) + (4\omega_0^2 + \lambda_0^2) \cdot \vec{u}_5^1(\vec{r}) = \frac{v_0^4 \cdot \lambda_0^4 \cdot \cos(\vec{k}\vec{r}) \cdot \vec{u}_0}{2^3 \cdot \upsilon^4 \cdot \omega_0^2 \cdot (3\lambda_0^2 - \omega_0^2) \cdot (3\lambda_0^2 - 3\omega_0^2)} \,, \tag{89}$$

$$\vec{u}_5^1(\vec{r}) = \frac{v_0^4 \cdot \lambda_0^4 \cdot \cos(\vec{k}\vec{r}) \cdot \vec{u}_0}{2^3 \cdot \upsilon^4 \cdot 4 \cdot \omega_0^4 \cdot (3\lambda_0^2 - \omega_0^2) \cdot (3\lambda_0^2 - 3\omega_0^2)} = \frac{v_0^4 \cdot \lambda_0^4 \cdot \cos(\vec{k}\vec{r}) \cdot \vec{u}_0}{2^4 \cdot \upsilon^4 \cdot 1 \cdot 2 \cdot \omega_0^4 \cdot (3\lambda_0^2 - \omega_0^2) \cdot (3\lambda_0^2 - 3\omega_0^2)} \,. \tag{90}$$

$$\Delta \vec{u}_6^1(\vec{r}) + (5\omega_0^2 + \lambda_0^2) \cdot \vec{u}_6^1(\vec{r}) = \frac{(\vec{v}(\vec{r}) \cdot \nabla) \cdot \vec{u}_5^1(\vec{r})}{\upsilon} \,. \tag{91}$$

$$\Delta \vec{u}_6^1(\vec{r}) + (5\omega_0^2 + \lambda_0^2) \cdot \vec{u}_6^1(\vec{r}) = -\frac{v_0^5 \cdot \lambda_0^5 \cdot \sin(2\vec{k}\vec{r}) \cdot \vec{u}_0}{2^5 \cdot \upsilon^5 \cdot 1 \cdot 2 \cdot \omega_0^4 \cdot (3\lambda_0^2 - \omega_0^2) \cdot (3\lambda_0^2 - 3\omega_0^2)} \,, \tag{92}$$

$$\vec{u}_6^1(\vec{r}) = \frac{v_0^5 \cdot \lambda_0^5 \cdot \sin(2\vec{k}\vec{r}) \cdot \vec{u}_0}{2^5 \cdot \upsilon^5 \cdot 1 \cdot 2 \cdot \omega_0^4 \cdot (3\lambda_0^2 - \omega_0^2) \cdot (3\lambda_0^2 - 3\omega_0^2) \cdot (3\lambda_0^2 - 5\omega_0^2)} \,. \tag{93}$$

$$\Delta \vec{u}_7^1(\vec{r}) + (6\omega_0^2 + \lambda_0^2) \cdot \vec{u}_7^1(\vec{r}) = \frac{(\vec{v}(\vec{r}) \cdot \nabla) \cdot \vec{u}_6^1(\vec{r})}{\upsilon} \,, \tag{94}$$

$$\Delta \vec{u}_7^1(\vec{r}) + (6\omega_0^2 + \lambda_0^2) \cdot \vec{u}_7^1(\vec{r}) = \frac{v_0^6 \cdot \lambda_0^6 \cdot \cos(\vec{k}\vec{r}) \cdot \vec{u}_0}{2^5 \cdot \upsilon^6 \cdot 1 \cdot 2 \cdot \omega_0^4 \cdot (3\lambda_0^2 - \omega_0^2) \cdot (3\lambda_0^2 - 3\omega_0^2) \cdot (3\lambda_0^2 - 5\omega_0^2)} \,, \tag{95}$$

$$\vec{u}_7^1(\vec{r}) = \frac{v_0^6 \cdot \lambda_0^6 \cdot \cos(\vec{k}\vec{r}) \cdot \vec{u}_0}{2^5 \cdot \upsilon^6 \cdot 1 \cdot 2 \cdot 6 \cdot \omega_0^6 \cdot (3\lambda_0^2 - \omega_0^2) \cdot (3\lambda_0^2 - 3\omega_0^2) \cdot (3\lambda_0^2 - 5\omega_0^2)} =$$
$$= \frac{v_0^6 \cdot \lambda_0^6 \cdot \cos(\vec{k}\vec{r}) \cdot \vec{u}_0}{2^6 \cdot \upsilon^6 \cdot 1 \cdot 2 \cdot 3 \cdot \omega_0^6 \cdot (3\lambda_0^2 - \omega_0^2) \cdot (3\lambda_0^2 - 3\omega_0^2) \cdot (3\lambda_0^2 - 5\omega_0^2)} \,. \tag{96}$$

At a result, we suppose that the resonance solution is

$$\vec{u}^{res}(t,\vec{r}) = \cos(\vec{k}\vec{r}) \cdot \vec{u}_0 \cdot \sum_{n=0}^{\infty} \frac{v_0^{2n} \cdot \lambda_0^{2n} \cdot \exp[-(2n \cdot \omega_0^2 + \lambda_0^2) \cdot \upsilon \cdot t]}{2^{2n} \cdot \upsilon^{2n} \cdot \omega_0^{2n} \cdot n! \prod_{m=1}^{n} (3\lambda_0^2 - (2m-1) \cdot \omega_0^2)} , \quad \frac{3\lambda_0^2}{\omega_0^2} \neq 2m-1 . \tag{97}$$

We define

$$A_n(t,v_0,\omega_0,\lambda_0) = \frac{v_0^{2n} \cdot \lambda_0^{2n} \cdot \exp[-(2n \cdot \omega_0^2 + \lambda_0^2) \cdot \upsilon \cdot t]}{2^{2n} \cdot \upsilon^{2n} \cdot \omega_0^{2n} \cdot n! \prod_{m=1}^{n} (3\lambda_0^2 - (2m-1) \cdot \omega_0^2)} , \tag{98}$$

$$A_n(v_0,\omega_0,\lambda_0) = \frac{v_0^{2n} \cdot \lambda_0^{2n}}{2^{2n} \cdot \upsilon^{2n} \cdot \omega_0^{2n} \cdot n! \prod_{m=1}^{n} (3\lambda_0^2 - (2m-1) \cdot \omega_0^2)} , \tag{99}$$

$$\vec{u}^{res}(t,\vec{r}) = \cos(\vec{k}\vec{r}) \cdot \vec{u}_0 \cdot \sum_{n=0}^{\infty} A_n(t,v_0,\omega_0,\lambda_0) = \cos(\vec{k}\vec{r}) \cdot \vec{u}_0 \cdot \sum_{n=0}^{\infty} A_n(v_0,\omega_0,\lambda_0) \cdot \exp[-(2n \cdot \omega_0^2 + \lambda_0^2) \cdot \upsilon \cdot t] . \tag{100}$$

$$\left| \frac{A_{n+1}(t,v_0,\omega_0,\lambda_0)}{A_n(t,v_0,\omega_0,\lambda_0)} \right| = \left| \frac{v_0^2 \cdot \lambda_0^2 \cdot \exp(-2 \cdot \omega_0^2 \cdot \upsilon \cdot t)}{2^2 \cdot \upsilon^2 \cdot \omega_0^2 \cdot (n+1) \cdot (3\lambda_0^2 - (2n+1) \cdot \omega_0^2)} \right| \to 0 \text{ for } n \to \infty, \; \forall t \in [0,\infty) . \tag{101}$$

According to d'Alembert's ratio test (Fihtengoltz, 2003b), the obtained series converges absolutely for any $t \in [0,\infty)$ if

$$\frac{3\lambda_0^2}{\omega_0^2} \neq 2m-1 . \tag{102}$$

Let's investigate the behavior of the solution if

$$\frac{3\lambda_0^2}{\omega_0^2} = 2m-1 . \tag{103}$$

For example, let's solve the system of Equation (69) if

$$\vec{k}_0 = \vec{k}, \; \lambda_0^2 = \vec{k}^2, \; \omega_0^2 = 3 \cdot \lambda_0^2, \; \vec{v}_0 \uparrow\uparrow \vec{k} . \tag{104}$$

We have the second equation as

$$\Delta\vec{u}_2(t,\vec{r}) - \frac{1}{\upsilon} \cdot \partial\vec{u}_2(t,\vec{r})/\partial t = \frac{v_0 \cdot k}{2 \cdot \upsilon} \cdot \exp(-4\lambda_0^2 \cdot t) \cdot \sin(2\vec{k}\vec{r}) \cdot \vec{u}_0 . \tag{105}$$

We find the solution as

$$\vec{u}_2(t,\vec{r}) = u_2 \cdot t \cdot \exp(-4\lambda_0^2 \cdot \upsilon \cdot t) \cdot \sin(2\vec{k}\vec{r}) \cdot \vec{u}_0 . \tag{106}$$

We obtain

$$-4 \cdot k^2 \cdot u_2 \cdot t - \frac{u_2}{\upsilon} + 4\lambda_0^2 \cdot u_2 \cdot t = \frac{v_0 \cdot k}{2 \cdot \upsilon} , \tag{107}$$

$$u_2 = -\frac{v_0 \cdot k}{2} . \tag{108}$$

$$\vec{u}_2(t,\vec{r}) = -\frac{v_0 \cdot \lambda_0 \cdot \vec{u}_0}{2} \cdot t \cdot \exp(-4\lambda_0^2 \cdot \upsilon \cdot t) \cdot \sin(2\vec{k}\vec{r}) . \tag{109}$$

The next equation is

$$\Delta\vec{u}_3(t,\vec{r}) - \frac{1}{\upsilon} \cdot \partial\vec{u}_3(t,\vec{r})/\partial t = -\frac{v_0^2 \cdot \lambda_0^2 \cdot \vec{u}_0}{2 \cdot \upsilon} \cdot t \cdot \exp(-7\lambda_0^2 \cdot \upsilon \cdot t) \cdot [\cos(\vec{k}\vec{r}) + \cos(3\vec{k}\vec{r})] . \tag{110}$$

We separate it into two equations:

$$\Delta\vec{u}_3(t,\vec{r}) - \frac{1}{\upsilon} \cdot \partial\vec{u}_3(t,\vec{r})/\partial t = -\frac{v_0^2 \cdot \lambda_0^2 \cdot \vec{u}_0}{2 \cdot \upsilon} \cdot t \cdot \exp(-7\lambda_0^2 \cdot \upsilon \cdot t) \cdot \cos(\vec{k}\vec{r}) \tag{111}$$

and

$$\Delta\vec{u}_3(t,\vec{r}) - \frac{1}{\upsilon} \cdot \partial\vec{u}_3(t,\vec{r})/\partial t = -\frac{v_0^2 \cdot \lambda_0^2 \cdot \vec{u}_0}{2 \cdot \upsilon} \cdot t \cdot \exp(-7\lambda_0^2 \cdot \upsilon \cdot t) \cdot \cos(3\vec{k}\vec{r}) . \tag{112}$$

We find the solution of the first equation as

$$\vec{u}_3(t,\vec{r}) = u_3 \cdot (t+a) \cdot \exp(-7\lambda_0^2 \cdot \upsilon \cdot t) \cdot \cos(\vec{k}\vec{r}) \cdot \vec{u}_0 \tag{113}$$

where $a$ is an unknown constant.
We obtain

$$(-k^2 + 7\lambda_0^2) \cdot u_3 \cdot t + (-k^2 + 7\lambda_0^2) \cdot u_3 \cdot a - \frac{u_3}{\upsilon} = -\frac{v_0^2 \cdot \lambda_0^2}{2 \cdot \upsilon} \cdot t \ , \tag{114}$$

$$u_3 = -\frac{v_0^2}{12 \cdot \upsilon} \ , \ a = \frac{1}{6 \cdot \upsilon \cdot \lambda_0^2} \ . \tag{115}$$

We find the solution of the second equation as

$$\vec{u}_3(t,\vec{r}) = u_3 \cdot (t+a) \cdot \exp(-7\lambda_0^2 \cdot \upsilon \cdot t) \cdot \cos(3\vec{k}\vec{r}) \cdot \vec{u}_0 \tag{116}$$

where $a$ is an unknown constant.

$$(-9k^2 + 7\lambda_0^2) \cdot u_3 \cdot t + (-9k^2 + 7\lambda_0^2) \cdot u_3 \cdot a - \frac{u_3}{\upsilon} = -\frac{v_0^2 \cdot \lambda_0^2}{2 \cdot \upsilon} \cdot t \ , \tag{117}$$

$$u_3 = \frac{v_0^2}{4 \cdot \upsilon} \ , \ a = -\frac{1}{2 \cdot \upsilon \cdot \lambda_0^2} \ . \tag{118}$$

We obtain the solution of Equation (110) as the sum of the solutions of Equations (111) and (112):

$$\vec{u}_3(t,\vec{r}) = -\frac{v_0^2}{12 \cdot \upsilon} \cdot (t + \frac{1}{6 \cdot \upsilon \cdot \lambda_0^2}) \cdot \exp(-7\lambda_0^2 \cdot \upsilon \cdot t) \cdot \cos(\vec{k}\vec{r}) \cdot \vec{u}_0 +$$

$$+ \frac{v_0^2}{4 \cdot \upsilon} \cdot (t - \frac{1}{2 \cdot \upsilon \cdot \lambda_0^2}) \cdot \exp(-7\lambda_0^2 \cdot \upsilon \cdot t) \cdot \cos(3\vec{k}\vec{r}) \cdot \vec{u}_0. \tag{119}$$

We have obtained the secular terms increasing the solution for some period of time. The time of growth (the blowup time) is not infinite due to the presence of the exponential function. This time is proportional to the value $\frac{1}{\upsilon \cdot \lambda_0^2}$ . It can be very large if $\upsilon \cdot \lambda_0^2 \ll 1$ and becomes infinitesimal if $\lambda_0^2 \to 0$ . The maximum of the time distribution of the solution approaches infinity.

## 3.5. Laminar flow without resonance

The increase of the solution as the result of the resonance of two fields is possible if a physical nature of the values $\vec{v}(t,\vec{r})$ and $\vec{u}(t,\vec{r})$ is different. For example, one is a fluid velocity, another is a magnetic field. If $\vec{v}(t,\vec{r})$ and $\vec{u}(t,\vec{r})$ describe a divergence-free and curl-free fluid velocity field the sum of them is a velocity of a fluid. It is considered that there exists no spontaneous growth of a velocity (kinetic energy) for a closed physical system in the absence of an external source (Landau, vol. 1, 1976). Actually, there is a change of basic values of solution parameters in comparison with their unperturbed values in higher approximations of nonlinear equations. This leads to the suppression of the resonance effect. We can remove the resonance terms using the method of Poincare (Landau & Lifshitz, 1976).

We suppose that

$$\vec{k} = \vec{k}^{(0)} + \vec{k}^{(1)} + \vec{k}^{(2)} + \ldots , \ (k^{(0)})^2 = \lambda_0^2, \ \vec{v}_0 \uparrow\uparrow \vec{k} \ , \ \vec{k}^{(0)} \uparrow\uparrow \vec{k}^{(1)} \uparrow\uparrow \vec{k}^{(2)} \uparrow\uparrow \ldots . \tag{120}$$

Our purpose is to delete the resonance term with $\cos(\vec{k}\vec{r})$ .
We modify the first equation of the system (72):

$$\Delta \vec{u}_1(\vec{r}) + \lambda_0^2 \cdot \vec{u}_1(\vec{r}) = 0 \ ,$$

$$\frac{\lambda_0^2}{k^2} \cdot \Delta \vec{u}_1(\vec{r}) + \lambda_0^2 \cdot \vec{u}_1(\vec{r}) = -(1 - \frac{\lambda_0^2}{k^2}) \cdot \Delta \vec{u}_1(\vec{r}) \ ,$$

$$\frac{\lambda_0^2}{k^2} \cdot \Delta \vec{u}_1(\vec{r}) + \lambda_0^2 \cdot \vec{u}_1(\vec{r}) = -\frac{(\vec{k}^{(0)} + \vec{k}^{(1)} + \vec{k}^{(2)} + \ldots)^2 - \lambda_0^2}{k^2} \cdot \Delta \vec{u}_1(\vec{r}) =$$

$$= -\frac{(k^{(1)})^2 + (k^{(2)})^2 + \ldots + 2 \cdot \lambda_0 \cdot k^{(1)} + 2 \cdot \lambda_0 \cdot k^{(2)} + 2 \cdot k^{(1)} \cdot k^{(2)} + \ldots}{k^2} \cdot \Delta \vec{u}_1(\vec{r}). \tag{121}$$

We construct the system using Equation (72) (Landau & Lifshitz, 1976):

$$\frac{\lambda_0^2}{k^2} \cdot \Delta \vec{u}_1(\vec{r}) + \lambda_0^2 \cdot \vec{u}_1(\vec{r}) = 0 \,,$$

$$\Delta \vec{u}_2(\vec{r}) + (\omega_0^2 + \lambda_0^2) \cdot \vec{u}_2(\vec{r}) = \frac{(\vec{v}(\vec{r}) \cdot \nabla) \cdot \vec{u}_1(\vec{r})}{\upsilon} - 2 \cdot \lambda_0 \cdot k^{(1)} \cdot \Delta \vec{u}_1(\vec{r}) \,,$$

$$\Delta \vec{u}_3(\vec{r}) + (2\omega_0^2 + \lambda_0^2) \cdot \vec{u}_3(\vec{r}) = \frac{(\vec{v}(\vec{r}) \cdot \nabla) \cdot \vec{u}_2(\vec{r})}{\upsilon} - \frac{(k^{(1)})^2 + 2 \cdot \lambda_0 \cdot k^{(2)}}{k^2} \cdot \Delta \vec{u}_1(\vec{r}), \tag{122}$$

…

$$\frac{\lambda_0^2}{k^2} \cdot \Delta \vec{u}_j(t,\vec{r}) + [(j-1) \cdot \omega_0^2 + \lambda_0^2] \cdot \vec{u}_j(\vec{r}) = \frac{(\vec{v}(\vec{r}) \cdot \nabla) \cdot \vec{u}_{j-1}(\vec{r})}{\upsilon} - \frac{(k^{(1)})^{j-1} + 2 \cdot \lambda_0 \cdot k^{(j-1)} + \ldots}{k^2} \cdot \Delta \vec{u}_1(t,\vec{r}),$$

….

The solution of the first equation is

$$\boxed{\phantom{xxxxxxxxx}}. \tag{123}$$

The second equation is

$$\Delta \vec{u}_2(\vec{r}) + (\omega_0^2 + \lambda_0^2) \cdot \vec{u}_2(\vec{r}) = -\frac{v_0 \cdot \lambda_0}{2 \cdot \upsilon} \cdot \sin(2\vec{k}\vec{r}) \cdot \vec{u}_0 - \frac{2 \cdot \lambda_0 \cdot k^{(1)}}{k^2} \cdot \Delta \vec{u}_1(\vec{r}). \tag{124}$$

There is no a term with $\cos(\vec{k}\vec{r})$ in the solution if $\vec{k}^{(1)} = 0$. We have the solution of the second equation

$$\vec{u}_2(\vec{r}) = \frac{v_0 \cdot \lambda_0 \cdot \vec{u}_0}{2 \cdot \upsilon \cdot (3\lambda_0^2 - \omega_0^2)} \cdot \sin(2\vec{k}\vec{r}) \,. \tag{125}$$

The third equation is

$$\Delta \vec{u}_3(\vec{r}) + (2\omega_0^2 + \lambda_0^2) \cdot \vec{u}_3(\vec{r}) = \frac{v_0^2 \cdot \lambda_0^2 \cdot \vec{u}_0}{2 \cdot \upsilon^2 \cdot (3\lambda_0^2 - \omega_0^2)} \cdot [\cos(\vec{k}\vec{r}) + \cos(3\vec{k}\vec{r})] - \frac{2 \cdot \lambda_0 \cdot k^{(2)}}{k^2} \cdot \Delta \vec{u}_1(\vec{r}). \tag{126}$$

$$\vec{u}_3(\vec{r}) = \frac{v_0^2 \cdot \lambda_0^2 \cdot \cos(3\vec{k}\vec{r}) \cdot \vec{u}_0}{4 \cdot \upsilon^2 \cdot (\omega_0^2 - 4\lambda_0^2) \cdot (3\lambda_0^2 - \omega_0^2)}, \text{ if } k^{(2)} = -\frac{v_0^2 \cdot \lambda_0}{4 \cdot \upsilon^2 \cdot (3\lambda_0^2 - \omega_0^2)} \,. \tag{127}$$

If $\omega_0^2 = \lambda_0^2$ we obtain

$$\boxed{\phantom{xxxxxxxxx}}, \tag{128}$$

$$\vec{u}_2(\vec{r}) = \frac{v_0 \cdot \vec{u}_0}{4 \cdot \upsilon \cdot \lambda_0} \cdot \sin(2\vec{k}\vec{r}) \,, \ \vec{k}^{(1)} = 0 \,, \tag{129}$$

$$\vec{u}_3(\vec{r}) = -\frac{v_0^2 \cdot \cos(3\vec{k}\vec{r}) \cdot \vec{u}_0}{24 \cdot \upsilon^2 \cdot \lambda_0^2} \,, \ k^{(2)} = -\frac{v_0^2}{8 \cdot \upsilon^2 \cdot \lambda_0} \,. \tag{130}$$

## 3.6. Laminar–turbulent transition

Let's consider the inhomogeneous equation

$$\partial \vec{u}(t,\vec{r})/\partial t + (\vec{v}(t,\vec{r}) \cdot \nabla)\vec{u}(t,\vec{r}) = \upsilon \cdot \Delta \vec{u}(t,\vec{r}) + \vec{f} \cdot \upsilon \cdot \exp(-\lambda_0^2 \cdot \upsilon \cdot t) \cdot \cos(\vec{\kappa} \cdot \vec{r}) \,, \tag{131}$$

$\vec{v}(t,\vec{r}) = \vec{v}_{POT}(t,\vec{r}) + \vec{v}_{SOL}(t,\vec{r})$ is the flow velocity,

$\vec{u}(t,\vec{r}) = \vec{v}_{SOL}(t,\vec{r})$ is the divergence-free flow velocity, and

$\vec{v}_{POT}(t,\vec{r})$ is the curl-free flow velocity.

We suppose that

$$\vec{v}(t,\vec{r}) = 2\upsilon \cdot \vec{\lambda} + \vec{V}(t,\vec{r}), \tag{132}$$

$\vec{\lambda}$ is the constant vector and $\vec{V}(t,\vec{r})$ is the flow velocity vector.

We consider the linearized equation

$$\Delta \vec{u}_1(t,\vec{r}) - \frac{1}{\upsilon} \partial \vec{u}_1(t,\vec{r}) / \partial t = 2 \cdot (\vec{\lambda} \cdot \nabla) \vec{u}_1(t,\vec{r}) + \vec{f} \cdot \exp(-\lambda_0^2 \cdot \upsilon \cdot t) \cdot \cos(\vec{\kappa} \cdot \vec{r}). \tag{133}$$

We perform the qualitative analysis of the resonance for this nonlinear equation (Landau & Lifshitz, 1976).

We find the solution of this equation as

$$\vec{u}_1(t,\vec{r}) = \vec{u}_0 \cdot \exp(-\lambda_0^2 \cdot \upsilon \cdot t) \cdot \cos(\vec{\kappa} \cdot \vec{r}). \tag{134}$$

We define

$$\vec{f}(t,\vec{r}) = \mathrm{Re}\{\vec{f} \cdot \exp[-\lambda_0^2 \cdot \upsilon \cdot t + i \cdot (\vec{\kappa} \cdot \vec{r})]\}, \tag{135}$$

$$\vec{u}_1(t,\vec{r}) = \mathrm{Re}\{\vec{u}_0 \cdot \exp[-\lambda_0^2 \cdot \upsilon \cdot t + i \cdot (\vec{\kappa} \cdot \vec{r})]\}. \tag{136}$$

We obtain

$$\vec{u}_0 = \frac{\vec{f}}{(\lambda_0^2 - \kappa^2) - 2i \cdot (\vec{\lambda} \cdot \vec{\kappa})} = \frac{\vec{f} \cdot \exp(i \cdot \delta)}{\sqrt{(\lambda_0^2 - \kappa^2)^2 + 4 \cdot (\vec{\lambda} \cdot \vec{\kappa})^2}}, \; tg\delta = \frac{2(\vec{\lambda} \cdot \vec{\kappa})}{\kappa^2 - \lambda_0^2}, \tag{137}$$

$$\vec{u}_1(t,\vec{r}) = \frac{\vec{f}}{\sqrt{(\lambda_0^2 - \kappa^2)^2 + 4 \cdot (\vec{\lambda} \cdot \vec{\kappa})^2}} \cdot \exp(-\lambda_0^2 \cdot \upsilon \cdot t) \cdot \cos(\vec{\kappa} \cdot \vec{r} + \delta). \tag{138}$$

We define the absolute value of the amplitude

$$u_0 = \frac{f}{\sqrt{(\lambda_0^2 - \kappa^2)^2 + 4 \cdot (\vec{\lambda} \cdot \vec{\kappa})^2}}. \tag{139}$$

We use the approximate expressions

$$\kappa^2 - \lambda_0^2 = (\kappa + \lambda_0) \cdot (\kappa - \lambda_0) \approx 2\lambda_0 \cdot \varepsilon \tag{140}$$

where $\varepsilon = \kappa - \lambda_0$,

$$2 \cdot (\vec{\lambda} \cdot \vec{\kappa}) \approx 2 \cdot (\lambda \cdot \lambda_0). \tag{141}$$

We obtain

$$u_0 = \frac{f}{2\lambda_0 \cdot \sqrt{\varepsilon^2 + \lambda^2}}. \tag{142}$$

$$u_0^2 = \frac{f^2}{4\lambda_0^2 \cdot (\varepsilon^2 + \lambda^2)}. \tag{143}$$

According to formulas (127) and (130), we accept that the absolute value of the wave vector $\vec{\kappa}$ is dependent on the amplitude $v_0$ due to the nonlinear effect. We define this dependence as (Landau & Lifshitz, 1976)

$$\lambda_0 + \chi \cdot v_0^2 \tag{144}$$

where $\chi$ is the function dependent on the anharmonic coefficients (see (127) and (130)).

It means that we replace the value $\lambda_0$ to $\lambda_0 + \chi \cdot v_0^2$ in $\kappa - \lambda_0$ for expression (143) (Landau & Lifshitz, 1976). As a result, we use the expression $\varepsilon - \chi \cdot v_0^2$ instead of $\varepsilon$ in (143).

We take into account that

$$\vec{v}(t,\vec{r}) = (\vec{u}_0 + \vec{v}_{POT}) \cdot \exp(-\lambda_0^2 \cdot \upsilon \cdot t) \cdot \cos(\vec{\kappa} \cdot \vec{r}), \; \vec{v}_0 = \vec{u}_0 + \vec{v}_{POT}, \tag{145}$$

where $\vec{v}_{POT}$ is the amplitude of the curl-free velocity, $\vec{u}_0 \perp \vec{v}_{POT}$, $\vec{v}_{POT} \parallel \vec{\kappa}$.

If the directions of $\vec{v}(t,\vec{r})$ and $\vec{\kappa}$ are defined we obtain that

$$u_0 = v_0 \cdot \sin\theta, \; \theta = const, \; 0 < \theta < \pi, \tag{146}$$

where $\theta$ is the constant angle between $\vec{v}(t,\vec{r})$ and $\vec{\kappa}$.

We obtain

$$u_0^2 \cdot [(\varepsilon - \chi_0 \cdot u_0^2)^2 + \lambda^2] = \frac{f^2}{4\lambda_0^2}, \ \chi_0 = \frac{\chi}{\sin^2\theta} \text{ or} \tag{147}$$

$$\varepsilon = \chi_0 \cdot u_0^2 \pm \sqrt{\left(\frac{f}{2\lambda_0 \cdot u_0}\right)^2 - \lambda^2}. \tag{148}$$

There is no difference between this expression and the formula presented in the work of (Landau & Lifshitz, 1976). Let's use the results of the analysis in this formula.

The greatest amplitude is

$$u_{0,\max} = \frac{f}{2\lambda_0 \cdot \lambda}. \tag{149}$$

If $f > f_k = \sqrt{\dfrac{32\lambda_0^2 \cdot \lambda^3}{3\sqrt{3}|\chi_0|}}$ \hfill (150)

we obtain the bifurcation solutions for the amplitude $u_0$ besides the unique one. So it is possible to describe laminar-turbulent transition and an opposite process. A flow separation is also considered. In addition, there are resonances at external forces whose wave vectors and frequencies are greatly different from their own ones (Landau & Lifshitz, 1976).

$$tg\,\delta = \frac{2(\vec{\lambda} \cdot \vec{\kappa})}{\kappa^2 - \lambda_0^2} \approx \frac{\lambda}{\varepsilon - \chi \cdot u_0^2}. \tag{151}$$

The phase shift between the external force and the velocity is dependent on the velocity amplitude and varies stochastically at the turbulence.

## 3.7. Equations for potential field and curl-free velocity

We use the equation

$$div(\vec{v}(t,\vec{r})) = \Delta\Phi(t,\vec{r}).$$

Its solution as a plane wave satisfies the Helmholtz equation

$$\Delta\Phi(t,\vec{r}) + \xi \cdot \Phi(t,\vec{r}) = 0, \tag{152}$$

$\Phi(t,\vec{r})$ is the scalar potential of the velocity vector field and $\xi$ is the value connected with the intermolecular interactions. In general, $\xi$ is dependent on space coordinates and can be complex-valued.

The equivalent system of the equations for the velocity is

$$\Delta\vec{v}_{POT}(t,\vec{r}) + \xi \cdot \vec{v}_{POT}(t,\vec{r}) = 0, \tag{153}$$

$$div(\vec{v}_{SOL}(t,\vec{r})) = 0. \tag{154}$$

These equations allow constructing the complete system of functions in space $R^3$.

We suppose that $\xi$ is a real value and

$$\xi > 0. \tag{155}$$

The solution of Equation (153) is

$$\vec{v}_{POT}(t,\vec{r}) = \vec{v}_{POT}(t) \cdot \cos(\vec{k}\vec{r} + \delta), \ k^2 = \xi, \ \delta \text{ is the phase shift.} \tag{156}$$

Let's verify the incompressibility condition

$$\left| div(\vec{v}(t,\vec{r})) \right| << \frac{U}{L} \tag{157}$$

for this solution.

$$\left| div(\vec{v}(t,\vec{r})) \right| = \left| (\vec{v}(t) \cdot \vec{k}) \cdot \sin(\vec{k}\vec{r}) \right| = \left| (v_{POT}(t) \cdot k) \cdot \sin(\vec{k}\vec{r}) \right|. \tag{158}$$

$$\left| \vec{v}_{POT}(t) \right| \sim U \, . \tag{159}$$

We obtain from the boundary conditions that

$$\cos(\vec{k} \cdot \vec{L}) = 0 \, , \tag{160}$$

$\vec{L} = (\pm L_x, \pm L_y, \pm L_z)$ is the boundary position.

We have that

$$\left| \vec{k} \right| \sim \frac{1}{L} \, . \tag{161}$$

We obtain from the scale definition that variations of $\left| \cos(\vec{k}\vec{r}) \right|$ are small if $\left| \vec{r} \right| << L$. Therefore,

$$\left| \vec{k}\vec{r} \right| << 1 \, . \tag{162}$$

$$\left| \sin(\vec{k}\vec{r}) \right| \approx \left| \vec{k}\vec{r} \right| \sim \frac{\left| \vec{r} \right|}{L} << 1 \, . \tag{163}$$

It means that condition (157) is performed. The fluid is incompressible.

We have also that

$$\xi \sim \frac{1}{L^2} \, . \tag{164}$$

It reflects the fact that the energy of the intermolecular interactions is extremely small.

It was shown in the previous part that laminar-turbulent transition is possible for a divergence-free velocity. We can construct equations which allow this for a potential of a curl-free field or a curl-free velocity.

Let's transform expression (152) to the equation similar to the one for oscillations of a mechanical system with an own frequency $\omega$ under the influence of a small nonlinear perturbation $\varepsilon \cdot F(t, \vec{r}, \Phi(t, \vec{r}), \partial \Phi(t, \vec{r})/\partial x, \partial \Phi(t, \vec{r})/\partial y, \partial \Phi(t, \vec{r})/\partial z)$ (Bogoliubov & Mitropolski, 1961):

$$\Delta(\Phi(t, \vec{r})) + \omega^2 \cdot \Phi(t, \vec{r}) = \varepsilon \cdot F(t, \vec{r}, \Phi(t, \vec{r}), \partial \Phi(t, \vec{r})/\partial x, \partial \Phi(t, \vec{r})/\partial y, \partial \Phi(t, \vec{r})/\partial z) \, , \tag{165}$$

where $\varepsilon$ is the small positive parameter and $F(t, \vec{r}, \Phi(t, \vec{r}), \partial \Phi(t, \vec{r})/\partial x, \partial \Phi(t, \vec{r})/\partial y, \partial \Phi(t, \vec{r})/\partial z)$ is the function which can be expanded to a Fourier series for the variable $\vec{r}$. The Fourier coefficients are polynomials in relation to $\Phi(t, \vec{r})$, $\partial \Phi(t, \vec{r})/\partial x$, $\partial \Phi(t, \vec{r})/\partial y$ and $\partial \Phi(t, \vec{r})/\partial z$. For instance, the equation is similar to the one for driven small nonlinear oscillations (Landau & Lifshitz, 1976) is

$$\Delta(\Phi(t, \vec{r})) + (\vec{\gamma} \cdot \nabla)\Phi(t, \vec{r}) + \vec{\omega}^2 \cdot \Phi(t, \vec{r}) = f(t) \cdot \cos(\vec{\omega}' \cdot \vec{r}) - \alpha \cdot \Phi(t, \vec{r})^2 - \beta \cdot \Phi(t, \vec{r})^3 \tag{166}$$

where $\vec{\gamma} = (\gamma_x, \gamma_y, \gamma_z)$ is the attenuation parameter, $\vec{\omega} = (\omega_x, \omega_y, \omega_z)$ are the own frequencies of system oscillations, $f(t)$ is the amplitude of the source of fluid motion which is presented by the periodic function with the frequency $\vec{\omega}' = (\omega_x', \omega_y', \omega_z')$, and $\alpha, \beta$ are the positive constants.

The similar equations can be derived for the curl-free velocity using expression (153).

The solutions of these equations describe the development of turbulence at a resonance as the equations for the divergence-free velocity (Landau & Lifshitz, 1976).
.

## 3.8. Magnetic Dynamo

We consider a mechanism of selfgenerating a magnetic field through movement of an electrically conducting fluid.

We take the equations of magnetohydrodynamics (MHD) for an incompressible fluid (Landau & Lifshitz, 1984) and modify them. The modified MHD equations look as follows

$$\rho \cdot \partial \vec{v}(t,\vec{r}) / \partial t + \rho \cdot (\vec{v}(t,\vec{r}) \cdot \nabla) \cdot \vec{v}(t,\vec{r}) = -\nabla \left( p(t,\vec{r}) + \frac{H^2}{8\pi} \right) + \eta \cdot \Delta \vec{v}(t,\vec{r}) +$$

$$+ \frac{1}{4\pi} (\vec{H}(t,\vec{r}) \cdot \nabla) \cdot \vec{H}(t,\vec{r}) + (\eta / 3 + \zeta) \cdot \nabla \, div(\vec{v}(t,\vec{r})), \tag{167}$$

$$\partial \vec{H}(t,\vec{r}) / \partial t = \frac{c^2}{4\pi\sigma} \Delta \vec{H}(t,\vec{r}) + (\vec{H}(t,\vec{r}) \cdot \nabla) \vec{v}(t,\vec{r}) - (\vec{v}(t,\vec{r}) \cdot \nabla) \vec{H}(t,\vec{r}) - \vec{H} \cdot div(\vec{v}(t,\vec{r})), \tag{168}$$

$$div(\vec{H}(t,\vec{r})) = 0, \tag{169}$$

$$div(\vec{v}(t,\vec{r})) = \Delta \Phi(t,\vec{r}), \tag{170}$$

$\vec{H}(t,\vec{r})$ is the magnetic field, $c$ is the speed of light, and $\sigma$ is electrical conductivity.

Equation (167) can be separated into parts the same way as Equation (48). As a result, we obtain the equations for the pressure as (54) and (55), and the equation for the velocity which is (see (56))

$$\rho \cdot \partial \vec{v}_{SOL}(t,\vec{r}) / \partial t + \rho \cdot (\vec{v}(t,\vec{r}) \cdot \nabla) \cdot \vec{v}_{SOL}(t,\vec{r}) = \eta \cdot \Delta \vec{v}_{SOL}(t,\vec{r}) + \vec{f}_{ex,SOL}(t,\vec{r}), \tag{171}$$

$$\vec{f}_{ex,SOL}(t,\vec{r}) = \frac{1}{4\pi} (\vec{H}(t,\vec{r}) \cdot \nabla) \cdot \vec{H}(t,\vec{r}). \tag{172}$$

All these equations can be solved by the way described in the previous part. The solution of Equation (168) obtains the resonant terms which are dependent on the velocity amplitude, $v_0^n$, $n = 1,2,3,\ldots$.

The time distribution grows to the maximum value and then vanishes exponentially. According to Equation (168), the increased magnetic field produces a high value of the amplitude of the fluid velocity as a result of resonance. It causes the next growth of the amplitude of the magnetic field. There exists a positive feedback between the moving current conducting fluid and the magnetic field. The growth of the velocity and magnetic field is accelerated, i.e. there is a blowup. The amplitude of the magnetic field can obtain very high values for a short period of time, even if the initial amplitude $u_0$ is comparatively small. As a result of strengthening, the amplitude of the magnetic field can exceed the threshold (150) and the turbulence is developed. The solutions of the equations have lost their uniqueness. The bifurcation occurs for the system parameters. The system is unstable and any fluctuation of the velocity parameters may cause flow detachment. The amplitude of the magnetic field falls down together with the amplitude of the fluid velocity. A new growth of the fluid velocity and magnetic field is originated from small stochastic perturbations in the conducting moving fluid (Landau & Lifshitz, 1984). The initial conditions of this growth define the generation of a magnetic excursion or reversal as a result.

## 4. Summary

It is possible to give a certain answer to the Clay Mathematics Institute's question. There exist no smooth functions $p(t,\vec{r})$, $\vec{v}(t,\vec{r})$ on $[0,\infty) \times R^3$ that satisfy the claims (A) or (B). For $\eta > 0$, the smooth solutions of the Navier-Stokes equations exist only in space of two variables.

We obtained the general solution of the mass continuity equation and studied it for the condition $\eta > 0$. The result is that there exists no smooth, divergence-free vector field $\vec{v}_0(\vec{r})$ on $R^3$ satisfying (C) or (D).

The answers to the questions formulated in the works of (Constantin, 2001; Ladyzhenskaya, 2003) are the following.

There are no conditions for smooth incompressible viscous velocities $\vec{v}(0,\vec{r})$ that ensure, in the absence of external input of energy, that the solutions of the Navier-Stokes equations exist for all positive time in three dimensions and are smooth.

The Navier-Stokes equations together with the initial and boundary conditions don't provide the deterministic description of the dynamics of an incompressible viscous fluid in space of three dimensions.

Under the conditions (A) and (B) the divergence-free solutions can be obtained only if the Navier-Stokes equation system takes the two-dimensional and linear form.

For obtaining the viscous solutions of the nonlinear problem it is necessary to break the strict divergence-free condition by introducing an implicit or explicit source in the right part of the mass continuity equation. Using this source, we apply a potential field whose Laplacian isn't equal to 0 at all points of $R^3$.

The exact solutions of the Navier-Stokes equations presented in the works of (Bertozzi & Majda, 2002; Landau & Lifshitz, 1987) are obtained using the linearization of the solutions or the infringement of the strict divergence-free condition (2) and/or by the introduction of a fictitious source in the right part of the mass continuity equation. In the last case, the solutions are obtained in curvilinear coordinates. The examples are a viscous fluid rotating by a disk (an infinite disk source), a fluid flow in a diffuser and confuser (an infinite linear source) and the Landau problem for a flooded jet (a point source) (Landau & Lifshitz, 1987).

However, we suppose that only a source geometrical factor is not enough for the description of nonlinear processes considering a viscous fluid flow. It is necessary to consider the molecular inelastic interactions in the mass continuity equation. These interactions change the intermolecular binding energy on the expense of the kinetic energy of a fluid. Their presence in the equation is carried out by the use of the curl-free potential velocity field. Due to this field the mass continuity equation can be reduced to the Helmholtz equation, whose solutions give a complete system of functions in $R^3$.

The modified equations describe an incompressible fluid flow. Due to the inelastic interactions an incompressible fluid sticks to a body surface. The analysis of the equation solutions shows the blowup of the laminar flow at the resonance, transition of the laminar flow to the turbulent one, and fluid detachment. This opens the way to solve the magnetic dynamo problem.

## Funding

The author received no direct funding for this research.

## Author details

A.V. Zhirkin[1]

E-mail: aleksej-zhirkin@yandex.ru

[1] Reactor Problem Laboratory, Department of Fusion Reactors, National Research Centre "Kurchatov Institute", Kurchatov Centre of Nuclear Technologies, Academician Kurchatov Square, 1, Moscow 123182, Russia.

Tel.: +7 499 196 70 41

Fax: +7 499 943 00 73

## Citation information